# Sea wave data reconstruction using micro-seismic measurements and machine learning methods

**Authors**: Lorenzo Iafolla[1], Emiliano Fiorenza[1], Massimo Chiappini[2], Cosmo Carmisciano[2], Valerio Antonio Iafolla[1]

[1]Assist in Gravitation and Instrumentation, via E. Stevenson n.3, CAP 00078, Monte Porzio Catone (Rome), Italy
[2] Istituto Nazionale di Geofisica e Vulcanologia, Via di Vigna Murata 605, 00143 Rome, Italy

**Correspondence:**
Lorenzo Iafolla
lorenzo.iafolla@outlook.it, lorenzo.iafolla@agi-tech.com

**Keywords**: Sea swell, ocean waves, sea state, sea wave period, buoy, machine learning.

## Abstract

Sea wave monitoring is key in many applications in oceanography such as the validation of weather and wave models. Conventional in situ solutions are based on moored buoys whose measurements are often recognized as a standard. However, being exposed to a harsh environment, they are not reliable, need frequent maintenance, and the datasets feature many gaps.

To overcome the previous limitations, we propose a system including a buoy, a micro-seismic measuring station, and a machine learning algorithm. The working principle is based on measuring the micro-seismic signals generated by the sea waves. Thus, the machine learning algorithm will be trained to reconstruct the missing buoy data from the micro-seismic data. As the micro-seismic station can be installed indoor, it assures high reliability while the machine learning algorithm provides accurate reconstruction of the missing buoy data.

In this work, we present the methods to process the data, develop and train the machine learning algorithm, and assess the reconstruction accuracy. As a case of study, we used experimental data collected in 2014 from the Northern Tyrrhenian Sea demonstrating that the data reconstruction can be done both for significant wave height and wave period.

The proposed approach was inspired from Data Science, whose methods were the foundation for the new solutions presented in this work. For example, estimating the period of the sea waves, often not discussed in previous works, was relatively simple with machine learning. In conclusion, the experimental results demonstrated that the new system can overcome the reliability issues of the buoy keeping the same accuracy.

# 1  Introduction

The complexity of the sea waves is mathematically described by the directional wave spectrum as a combination of waves propagating in different directions with different wavelengths (Talley et al. 2011). The knowledge of the directional wave spectrum is key in several applications such as coastal management and design of coastal and offshore structures (e.g., ports and renewable energy platforms). Indeed, forces on piles, breakwaters, offshore structures as well as wave-induced coastal erosion, all depend on the directional wave spectrum. Recently, accurate wave measurements are required also in marine renewable energy industry for engineering design, and for resource and performance assessments (Thies et al. 2014). Key applications concern verification and data assimilation into weather and sea waves models to improve their accuracy (Krogstad et al. 2005; Mentaschi et al. 2015).

Fostered by the interest on the numerous applications, technology for ocean observation and monitoring has made significant advances in the last decades (Ardhuin et al. 2019; Lin and Yang 2020). For example, if in the middle of the twentieth century the measurements of the directional wave spectrum were a major achievement, nowadays many systems based on different measuring principles are affordable for operational use. In (Krogstad et al. 2005; Souza et al. 2011), good reviews of the available sea state monitoring systems are provided distinguishing two families: remote sensing and in situ. Examples of remote sensing systems are those based on radars. These can be ground-based (Wyatt et al. 2003; Lopez and Conley 2019; Novi et al. 2020), ship-based (Izquierdo et al. 2004), airborne (Voronovich and Zavorotny 2017; Le Merle et al. 2019; Sun et al. 2020), as well as spaceborne (Macklin and Cordey 1991; Aouf et al. 2021) and rely on the analysis of the backscattered intensity and/or the Doppler spectrum of radar signals. Examples of in situ systems are the subsurface devices, such as pressure and acoustic sensors, but the most common are those based on moored buoys instrumented with motion sensors such as accelerometers, gyroscopes, or GPS as described in (Herbers et al. 2012; Andrews and Peach 2019) and in Datawell website[1]. The directional wave spectrum is calculated from the raw measurements by using algorithms based on the hydrodynamics characteristics of the hull.

Buoy technology is well established and recognized as a standard since decades, however, uncertainties have been well demonstrated as discussed in (Ashton and Johanning 2015; Ardhuin et al. 2019; Jensen et al. 2021). For example, some issues might arise from the mooring (Niclasen and Simonsen 2007) or from biofouling (Campos et al. 2021). Furthermore, buoys are installed in a harsh environment, at the mercy of sea waves, wind, storms, and other possible causes of damage. For instance, they might be accidentally damaged by ships when moored next to naval routes, e.g., close to a port. Therefore, buoys are vulnerable to system failures, communication problems, breakage of the mooring, vandalism, etc., and require continuous maintenance. Consequently, data gaps might be very large and frequent, while maintenance costs might be very high. For example, in (Picone 2009) the analysis of the data collected from 2002 to 2006 by the 14 buoys of the Italian Data Buoy Network (Piscopia et al. 2003; Bencivenga et al. 2012) revealed that the missing data of the most reliable buoy (Cetraro) were the 15.5% of the total, whereas those of the worst reliable (P. d. Maestra) were the 88.8% (performance of other buoys are shown in Figure 1).

The maintenance issues are not specific of the Italian Data Buoy Network, and many works focus on missing data reconstruction using machine learning (ML) and data science methods. For example in (Vieira et al. 2020), a method based on artificial neural networks is presented to fill the waves record gaps using offshore hindcast and wind information. In (Jörges et al. 2021), a Long Short-Term Memory neural network was used to reconstruct the significant wave height from sea state time series, weather data of adjacent buoys, and bathymetric data. In (Agrawal and Deo 2002), a first order "auto regressive integrated moving average" (ARIMA) model, e.g., see (Nielsen 2019), was used and compared to a feed-forward neural-network for making sea wave predictions.

---

[1] https://www.datawell.nl/Products/Buoys.aspx

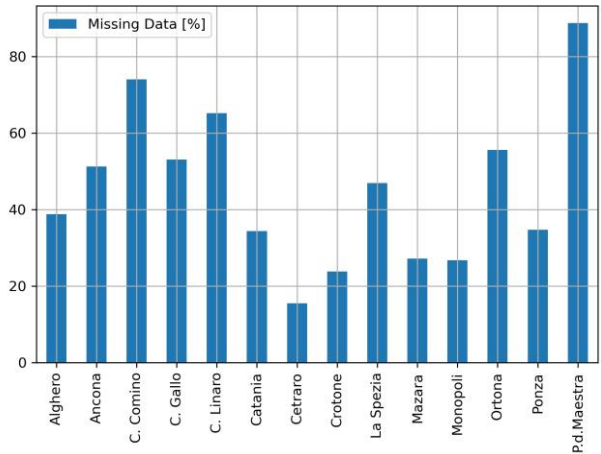

*Figure 1. Missing data percentage of the 14 buoys of the Italian Data Buoy Network from 2002 to 2006. Source (Picone, 2009).*

Besides data reconstruction, ML is key in many methods for oceanography (Lou et al. 2021). In fact, ML algorithms can find a mathematical model (called ML model) which produces the desired output when applied to a set of input data (called "training data"). The interesting part is the prediction, in which the ML model generates the correct outputs also when applied to new input data, distinct from the training data, coming from the same source. An important differentiation is between supervised learning and unsupervised learning. In the first case, the desired output is available for the training data, in the latter case, the desired output is not available. Supervised learning is more appropriate for missing data reconstruction because the desired outputs are typically available. For example, the desired output might be the sea wave data measured by a buoy whereas the input data comes from adjacent buoys, wind measurements, offshore hindcast, or, as we do in this work, micro-seismic data. During the regular operation, all data, including the desired output, are available and can be used for training. Instead, when data from the buoy are missing, the ML model will predict them from the input data. There are several ML algorithms able to make both regression (predict a continuous value, such as the significant wave height) and classification (predict a class, such as the degrees of the Douglas Sea scale). For example, there is whole family of algorithms, called artificial neural networks, inspired by natural neural networks. Other algorithms are the results of mathematical approaches, such as linear models, support vector machines, and decision trees. An introduction to machine learning is in (Burkov 2019), while a more comprehensive exposition is in (Géron 2019).

In this work, we will use micro-seismic data acquired from an onshore seismometer as input data. Actually, it is well known that sea waves are source of a micro-seismic signal which is detectable from onshore, even at many kilometers from the coast. Although this phenomenon has been discovered more than a century ago, the first geophysical model was presented in (Longuet-Higgins 1950); more details are provided in Section 2. Improvements of this model have been proposed. For example, in (Ardhuin et al. 2011; Ardhuin et al. 2012) three different geophysical models were introduced for three different types of events. In other works, the relationship between the micro-seismic signal and the sea waves has been investigated with focus on specific locations (Barruol et al. 2006; Davy et al. 2016; Ferretti et al. 2018) and on specific events (Cutroneo et al. 2021). In (Ferretti et al. 2013), an algorithm based on Markov chain Monte Carlo is used to determine the model parameters for a study conducted in the Ligurian coast (Italy). In (Barruol et al. 2016), the authors evaluated the correlation between the polarization of the micro-seismic signal and the swell propagation direction. In (Serafino et al. 2021), simultaneous measurements of a micro-seismic based system (called OS-IS) and those of a radar system were compared for the first time. In (Cannata et al. 2020), a machine learning method (specifically, a random forest) was proposed to reconstruct the spatial distribution of sea wave height, as provided by hindcast maps of sea wave models, by using micro-seismic data from multiple seismic stations. In (Moschella et al. 2020), a network of broadband seismic stations was used to investigate the micro-seismic signals from Ionian and Tyrrhenian Sea and, importantly, it was demonstrated that the signal detected by seismic stations closer to the sea contain more information concerning the sea state than the others.

As demonstrated in (Iafolla et al. 2015), systems based on micro-seismic measurements outperform the systems based on moored buoys in terms of reliability and sustainability. Consequently in this work, we propose a measuring system consisting of a moored buoy, a micro-seismic measuring station, and a supervised machine learning algorithm to provide accurate sea wave measurements (specifically, significant waves height $H_S$, peak period $T_p$, and mean period $T_m$) continuously and reliably. This measuring system will typically provide the measurements from the buoy, which we will use as the desired output for the training data. Instead, the micro-seismic data are the input data, processed by the machine learning algorithms to reconstruct (i.e., "predict" in ML jargon) the sea wave data and fill the gaps due to the failures of the buoy. Therefore, the proposed system features the accuracy of the buoys and the reliability of the micro-seismic method overcoming the limitations of the two methods taken separately. In this work, we present the methods to preprocess the data, develop the ML models, and evaluate their accuracy. As a case of study, we will use the data recorded simultaneously by a buoy and a micro-seismic based system to validate the proposed methods and to assess their accuracy.

## 2 Background – From micro-seismic signals to sea waves parameters

In this section, we introduce a simple model, based on the Longuet-Higgins's one, to derive the sea waves parameters from the micro-seismic signals. This model provides basic notions and it was inspirational to develop the ML methods. Furthermore in this work, it was used as a benchmark for comparisons.

Longuet-Higgins showed that a peak of the micro-seismic spectrum is related to sea waves travelling in opposite direction (e.g., waves generated by coastal reflection) with similar frequencies. A peculiarity is that the peak of the micro-seismic spectrum has doubled frequency compared to that of the sea waves. Another phenomenon, with lower seismic energy and same frequencies as the sea waves, originates by the interaction of the waves with a sloping bottom. A recent review and detailed description of these phenomena are given in (Ardhuin et al. 2019). Considering the Longuet-Higgins phenomenon, a simple mathematical equation to calculate the significant wave height $H_S$ from the micro-seismic power spectral density $S(f)$ (with $f$ being the frequency) is the following (Bromirski et al. 1999).

$$H_S = \alpha \cdot \sqrt{\int_{f_{min}}^{f_{max}} S(f) \cdot df} + \beta \qquad (1)$$

In the previous equation, $\alpha$ and $\beta$ are parameters of the model. The limits of integration $f_{max}$ and $f_{min}$ are also parameters and define a bandwidth that must contain all the micro-seismic signal generated by the sea waves. However, this bandwidth should not be too large, in order to avoid extraneous micro-seismic contributions that would worsen the accuracy of the evaluation of $H_S$. To calculate $T_m$ and $T_p$, one can simply use $S(f)$ in place of the sea wave power spectral density. However, we recall that the frequency of the micro-seismic signal is doubled compared to that of the sea waves. Therefore, $T_p$ is $2/f_p$ (with $f_p$ being the peak frequency of $S(f)$) and $T_m$ is defined by the following mathematical formula (Krogstad et al. 1999).

$$T_m = 2 \cdot \frac{\sqrt{\int_{f_{min}}^{f_{max}} S(f) \cdot df}}{\sqrt{\int_{f_{min}}^{f_{max}} f \cdot S(f) \cdot df}} \qquad (2)$$

# 3    Materials and methods

## 3.1   Measuring systems

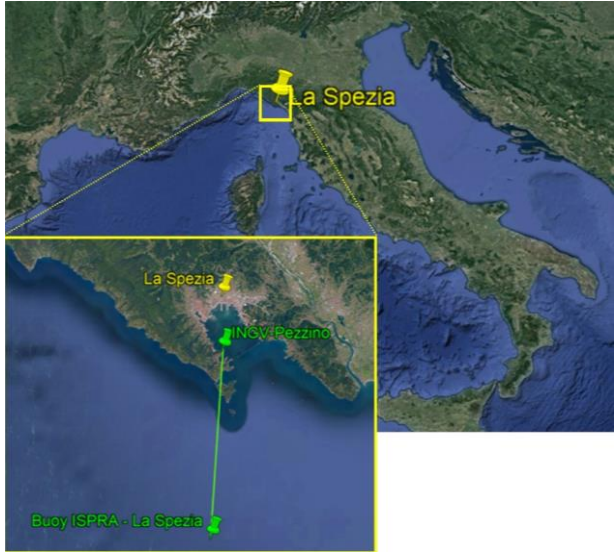

*Figure 2. Locations of La Spezia city (yellow marker), OS-IS station (INGV-Pezzino, green marker), and buoy ISPRA-La Spezia (green marker). The distance, indicated by the green line, between OS-IS and the buoy is about 16 km.*

In this work, we used data collected in 2014 by a sea wave monitoring system, called OS-IS (Ocean Seismic – Integrated Solution), based on the micro-seismic method and data by the buoy of the Italian Data Buoy Network moored in proximity of La Spezia. The latter data are publicly available on the European Marine Observation and Data Network (EMODnet)[2] and a detailed description of the Italian Data Buoy Network is in (Bencivenga et al. 2012). The buoy ISPRA-La Spezia was moored at latitude 43°55'45.00"N and longitude 9°49'40.00"E (see Figure 2). At about 16 km (green line), the OS-IS station was installed in the basement of Villa Pezzino, the INGV (National Institute for Geophyisics and Volcanology) labs of Porto Venere (La Spezia), at latitude 44°4'24.19"N and longitude 9°50'22.84"E.

The OS-IS station at INGV-Pezzino was installed in December 2013 in the framework of a project called Wind, Ports, and Sea (Bonino et al. 2015) funded by the European Cross-border Programme "Italy–France Maritime 2007-2013". A detailed description of OS-IS is provided in (Iafolla et al. 2014; Iafolla et al. 2015; Carmisciano et al. 2016) and its simplified schematic is shown in Figure 3. The core is the high-sensitivity three-axial accelerometer developed by AGI srl (Figure 4). Its background noise level is lower than $10^{-7}$ m/s$^2$/√Hz in the bandwidth of interest for measuring the micro-seismic signal generated by the sea waves (from ~$4·10^{-2}$ Hz to ~1 Hz). The sampling rate of the accelerometer was set to 10 Hz, which is about 10 times bigger than the highest frequency of interest. Although all three components (x, y, and z) of the acceleration were available, in this work, we used only the vertical component, aligned to the local gravity.

The measurements from the accelerometer were transmitted, through the internet, to a server for data storage and processing. Further descriptions regarding the computing system are reported in Section 1 of supplementary material of this paper.

## 3.2  Graphical tools and validation metrics for data analysis

In this work, data analysis was key for two main tasks. The first was to explore data, identify anomalies, and, consequently, remove noisy records. The second was to validate the ML models and assess their performance. The validation is done by comparing two datasets: the predicted values and the desired values. For example, we compared the $H_s$ values predicted by ML models and the $H_s$ values measured by the buoy.

Performing the former tasks require several tools, both graphical and numerical. We used well-known graphical tools, such as time plots, scatter plots, and histograms, as well as less used (at least in -sea wave to micro-seismic- data analysis) tools such as empirical cumulative distribution function (ECDF) plots and hexagonal binning plots.

---

[2] https://www.emodnet-physics.eu/map/platinfo/piroosplot.aspx?platformid=8712

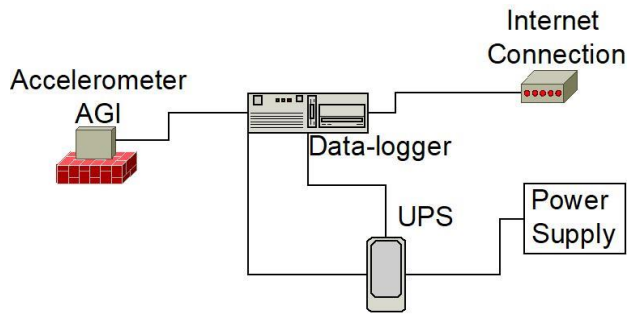
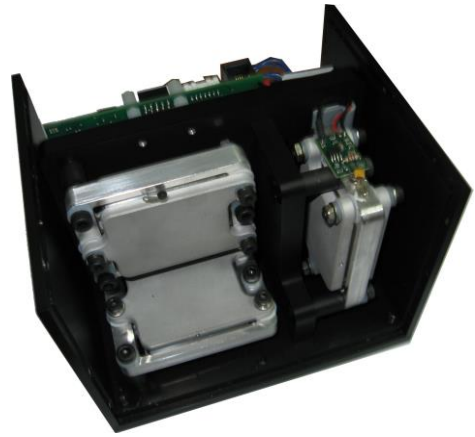

*Figure 3. Simplified schematic of an OS-IS station. The UPS (Uninterruptible Power Supply) improves the reliability of the system. All instrumentation is installed indoor, sheltered from the environment (rain, humidity, dust, etc.). One OS-IS station might also include a weather station and a GPS receiver for weather monitoring and precise timing.*

*Figure 4. Picture of a three-axis accelerometer by AGI srl. The three grey elements are transducers for measuring the x, y, and z components of the acceleration. Typically, the accelerometer is closed to protect the transducers and the acquisition electronics.*

In **ECDF plots** (Downey 2014), the value of the ECDF (y-axis) at any specified point $x_0$ of the x-axis is the fraction of observations of the variable x that are smaller than or equal to $x_0$. In formula, $ECDF(x_0)=P(x \leq x_0)$, where $P(x \leq x_0)$ is the probability that $x \leq x_0$. ECDFs are useful for comparing sea wave distributions (Krogstad et al. 1999) because they smooth out random variations which, instead, are typical in histograms. Furthermore, synchronization of the records is not necessary, as with time plots, because the timing is lost in ECDFs.

**Hexagonal binning plots** are an alternative to the more common scatter plots to show and verify the relationship between two variables (Bruce et al. 2020). The records are grouped into hexagonal bins whose color indicates the number of records in that bin. The hexagonal binning plots are ideal to display large datasets that would appear as a monolithic cloud of points in conventional scatter plots. To make hexagonal binning plots, we used the *hexbin* function of the Matplotlib library of Python.

Other common methods for comparing two datasets are the validation metrics such as **RMSE** (Root Mean Squared Error), **MAE** (Mean Absolute Error), and **Pearson correlation coefficient** (Géron 2019; Bruce et al. 2020). To compute them, we used the functions *mean_squared_error*, *mean_absolute_error* (from sklearn.metrics library of Python) and *corrcoef* (from numpy library).

## 3.3 Data pre-processing and cleaning

**Exploratory data analysis** (EDA) is a process for gaining an insight into a dataset (Downey 2014). A golden rule in ML is to always perform an EDA before preparing the data and feeding them into a ML algorithm. For example in this study, it is relevant to know how the values of $H_s$, $T_p$, and $T_m$ are distributed and if there is any relationship between them. Section 2 of the supplementary material reports the results of an EDA conducted on the dataset used in this study.

**Data augmentation.** The amount and the quality of the training data is important for successfully training ML algorithms. In other words, a large dataset well representing the population of possible inputs is desirable. Sometimes, new training data can be generated from the available ones, this is called "data augmentation" in ML jargon. For example, it might be sufficient to flip one image about its central axis to obtain a new sample image for training. However, the new sample should not be too much alike the original one, otherwise it will not determine any improvement of the training.

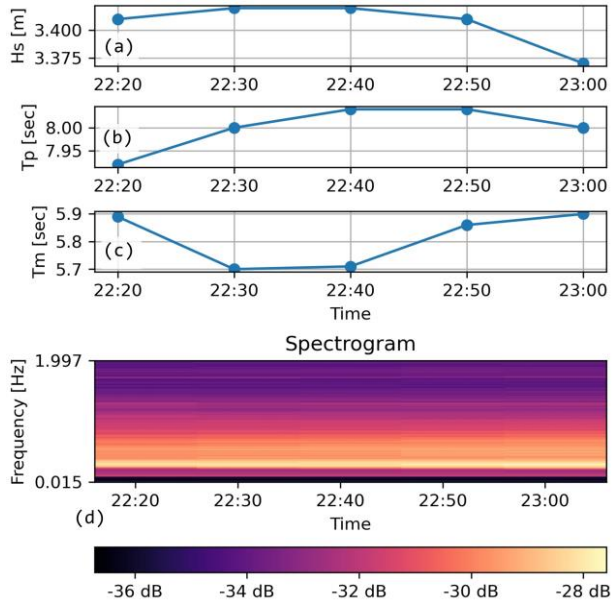

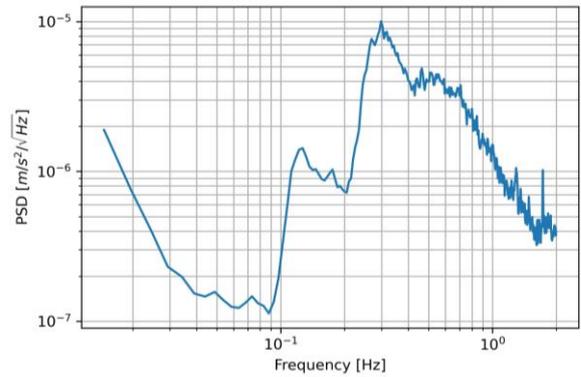

*Figure 6. Power spectral density (PSD) of the micro-seismic signal recorded on the 5th February 2014 at 23:00. The corresponding sea waves parameters from the buoy were: $H_s$=3.37 m; $T_p$=8.00 s; and $T_m$=5.90 s.*

*Figure 5. The time plots (a, b, and c) show five buoy records from 22:20 to 23:00 on the 5th February 2014. In (d) is shown the spectrogram of the micro-seismic signal, i.e., the five columns represent the five PSDs corresponding to the timings indicated by the labels on the x axis.*

In our work, the size of the dataset was limited by the sampling period of the buoy, which was 30 minutes. To increase this size, we interpolated the buoy data to a sampling period of 10 minutes. To do so, we used the *interpolate* method of the Pandas library of Python with the interpolation parameter set to "quadratic". We did not further decrease the sampling period to avoid feeding the ML algorithms with samples too much alike each other. For example, records of the sea wave parameters are displayed in the time plots of Figure 5 (a, b, and c). In such short timing, the variations are already quite small because of the inertia of the sea state. Consequently, further interpolation would not determine any improvement of the training because the newly generated records would be almost identical to the existing ones.

**Feature engineering.** Data pre-processing is called "feature engineering" in ML jargon, where the features are the input variables of the algorithm: for example, if the input is an array, each of its elements is a feature. In other words, feature engineering defines the shape of the ML algorithm input and impacts significantly its performance (e.g., the accuracy). Inspired by the model introduced in Section 2 and timeseries forecasting methods (Nielsen 2019), we defined and tested two methods to engineer the features, i.e., to pre-process the micro-seismic data. Both these methods use the power spectral density (PSD) of the micro-seismic signal, like that shown in Figure 6. Details on the method to compute the PSD are provided in Section 3 of supplementary material.

We defined the first feature engineering method inspired by Equations (1) and (2), which contain the micro-seismic spectrum $S(f)$. This method merely consists in using the PSD as ML algorithm input.

We defined the second feature engineering method inspired by the analysis of the properties (stationarity, autocorrelation, partial-autocorrelation, etc.) of the timeseries $H_s$, $T_p$, and $T_m$ and, specifically, by the possibility to predict their present value from their past values (Nielsen 2019). For example, Figure 7 shows that an ARIMA(3,1,5) model can forecast $H_s$ using its past five values. This suggests that past $H_s$ values carry useful information to evaluate the present value. Similarly, we supposed that the past records of the PSD might carry useful information to evaluate the present sea state. To leverage this, we used a spectrogram, i.e., an image combining the last five PSDs like that shown in Figure 5 (d), as ML algorithm input. Similar spectrograms with more than five PSDs could be used, however they require more memory and computational power. More details on the computation of the spectrogram are provided in Section 3 of the supplementary material.

Summarizing, the first feature engineering method consists in feeding the ML algorithm with the present PSD of the seismic signal; the second in feeding it with the last five PSDs (i.e., the spectrogram).

**Data cleaning** is aimed at achieving optimal results in ML and at checking the quality of the records during the operational use. Given the big amount of data and the real-time applications, it is also important to automatize the cleaning and the quality checking.

Data from the buoy were already classified with a quality check label ("good value", "bad value", etc.). We merely discarded all those samples whose label was not "good value". Other records were dropped because the corresponding micro-seismic data were not available. The number of available and missing records is reported in Suppl. Table 1. In addition, we picked just the variables of our interest (i.e., timestamp, $H_s$, $T_p$, $T_m$, and swell direction) and dropped all other variables (e.g., temperature of the water, intensity of the wind).

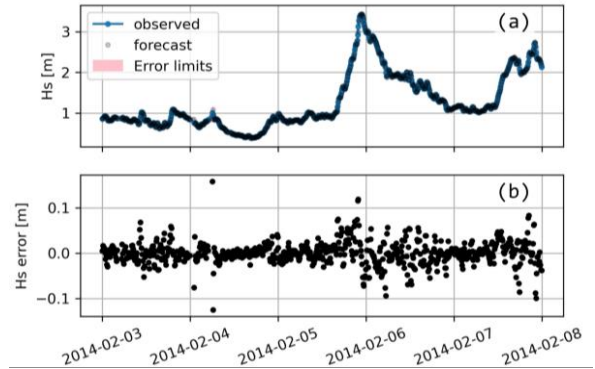

*Figure 7. (a) $H_s$ observed by the buoy and forecasted with an ARIMA(3,1,5) model. Each forecast is evaluated using the past observed values. The error limits are too small to be visible on this scale. (b). Deviation of the forecasted values from the observed ones.*

To perform micro-seismic data cleaning, it was useful to compute a timeseries $I_{PSD}(t)$, or simply $I_{PSD}$, of the mean values of the elements of the PSDs. For example, $I_{PSD}$(5$^{th}$ February 2014 at 23:00) is the mean value of the points shown in the plot of Figure 6. For convenience, we standardized $I_{PSD}(t)$ by subtracting its mean value and dividing by its standard deviation. Furthermore, it was useful to compute the timeseries, diff($I_{PSD}$), of the variations of each record of $I_{PSD}$ with respect to the recent previous ones. To do so, we used a rolling window (4 hours wide) technique over $I_{PSD}$; i.e., diff($I_{PSD}$) is the difference between the value of last point and the mean value within the window.

Using $I_{PSD}(t)$ and diff($I_{PSD}$) we could identify noisy data caused by micro-seismic disturbances. Most common examples of the latter are earthquakes and human activities, such as people walking next to the accelerometer. These disturbances might last for several minutes and they might affect several consecutive input samples (i.e., PSD records). This is visible in the time plots of $I_{PSD}$, such that in Figure 8 (a). Bunches of points, highlighted in red, are clearly displaced away from the others, indicating an anomaly. Sometimes, these anomalies are very high and setting a threshold over $I_{PSD}$ is sufficient to spot them automatically. An example is the bunch of points recorded on the 24$^{th}$ of May. However, their values are often lower than peaks due to the sea waves. For example, the bunch of red points recorded on the 9$^{th}$ of May are clearly anomalies because they do not have correspondence with the $H_s$ measured by the buoy (Figure 8 (c)). Still, their $I_{PSD}$ values are lower than the peak on the 13$^{th}$, which is due to sea waves. To automatically identify these points, we used diff($I_{PSD}$), shown in Figure 8 (b), where these points emerge from the rest.

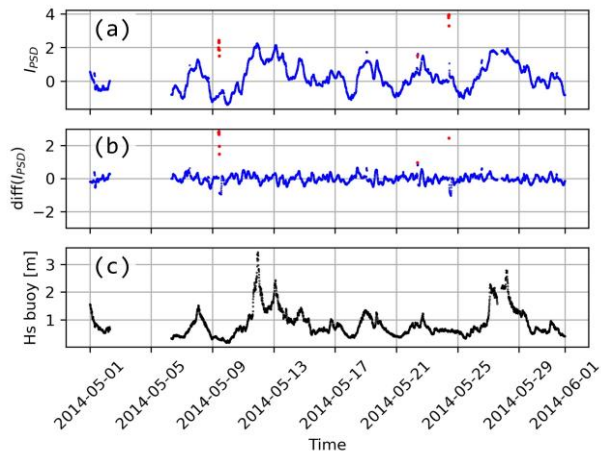

*Figure 8. Time plots of (a) $I_{PSD}$, (b) diff($I_{PSD}$) (as defined in Section 3.3), and (c) $H_s$. Anomalies are in red.*

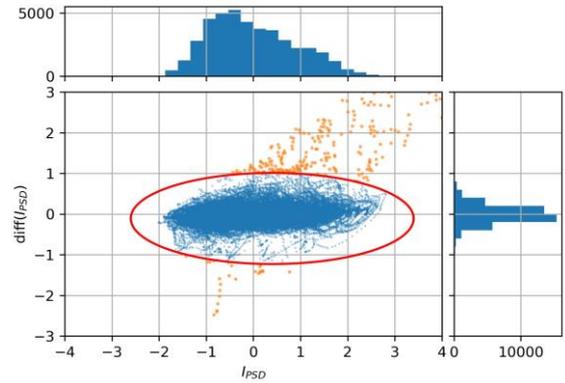

*Figure 9. Scatter plot and corresponding histograms of $I_{PSD}$ vs diff($I_{PSD}$). All points outside the ellipse, colored in orange, are identified as anomalies.*

To automatically identify and discard anomalies, we made the scatter plot of $I_{PSD}$ vs diff($I_{PSD}$) as shown in Figure 9. Most of the records are close to the center (0,0), as also visible from the side and top histograms, whereas the anomalies are far away. Consequently, we defined an ellipse (shown in red), inside which all good records are expected to be. The anomalies, colored in orange, are outside the ellipse and, consequently, easy to discard. Defining the size and the position of the ellipse is trade-off: when it is too small, some good records might be wrongly discarded; when it is too large, some anomalies might not be identified as such. In our case, we defined the $I_{PSD}$ axis length equal to 6 and the diff($I_{PSD}$) axis length equal to 2.2.

## 3.4 ML algorithms and models validation

The **ML algorithms** able to predict the sea state with the highest accuracy were ensembles of decision trees (bagging, boosting, and random forests) and convolutional neural networks (a special case of deep learning algorithms).

A decision tree (DT) (Géron 2019; Bruce et al. 2020) is a process that sequentially examines, one per time, the features of an input sample assigning it to an output value. However, only a finite number of possible output values is defined during the training, consequently, the output is discrete.

Ensemble methods overcome the limit of the discrete output and improves the accuracy of the DTs. In general, an ensemble is a group of predictors, e.g., a DT, making predictions on the same input sample. The output of the ensemble is an aggregation, e.g., the average, of all predictions. There are several ensemble methods; we used bagging, boosting, and random forest (Random F.), which are among the most commonly used (Géron 2019). We implemented the ensembles using the dedicated functions of the Scikit Learn library of Python and we fed them with the PSDs defined in Section 3.3. The most important parameters of the ensemble methods are the max depth of the DTs and the number of predictors (called n_estimators in Scikit Learn). We used max depth equal to 7 and number of predictors equal to 15 or 20.

Finally, we trained and tested convolutional neural networks (CNNs), which were developed for processing images such as the spectrograms described in Section 3.3. For example, CNNs are famous for being able to classify pictures of dogs, cats, or other objects. Adjusting their output layer, they can also make regression as needed for the sea wave variables. A CNN consists of several layers of different types that are stacked and set accordingly to the task (Géron 2019). Very complex tasks usually require high number of layers, large training datasets, and the training takes very long time (or very large computational power). The configurations we tested in this work are reported in Suppl. Table 2 of supplementary material.

**ML models validation**. After a ML algorithm has been trained using training data, it provides a ML model. This can predict the output given only the input data; however, its performance should be assessed before using it. To make this assessment, the graphical methods and the metrics introduced in Section 3.2 can be used to compare the outputs of the model with the desired outputs. This assessment is called validation and it is performed over a validation dataset that must contain also the desired outputs. However, a ML algorithm might be able to memorize (this is called overfitting) the training data without being able to generalize over new input data. Therefore, keeping the training data separated from the validation data is key to validate correctly the ML model even when it is overfitting. One method is based on holding out a validation subset and training the ML algorithm by using the rest of the dataset. However, the validation results might depend on the chosen validation subset. To overcome this issue, we used a method called cross-validation (Géron 2019; Bruce et al. 2020); the idea is to split the full dataset in subsets, e.g., 12 (see Figure 10). The ML algorithm is then trained 12 times, each time holding out a different subset and using it only for validation. Merging the results, we can obtain validation estimates over the full dataset.

The bootstrap method (Bruce et al. 2020) can be used to assess the confidence intervals of the validation metrics. In this work, we used a similar method that allows us to inspect the performance of the ML model over the time. The idea is to estimate the validation metrics over multiple (e.g., 50) consecutive time intervals. Specifically, first we divided the validation dataset in 50 consecutive subsets, then we estimated the validation metrics of each subset to obtain a timeseries. This can be plotted to see how the ML model behaved over time and its ECDF can be used to easily assess the confidence intervals. Examples are shown in Section 4, Figure 11.

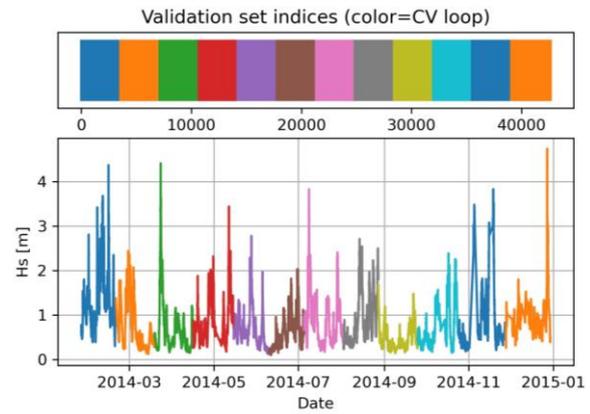

*Figure 10. In cross-validation, the full dataset is split in subsets (e.g.,12) of the same size. In this figure, different colors indicate different validation subsets.*

# 4 Results

The dataset used for training and validating the ML models was made of 42603 simultaneous records, whereas 9957 records were not available or discarded during the data cleaning (see Suppl. Table 1 for more details).

The validation metrics of the best performing ML models and those of the conventional model introduced in Section 2 (indicated as OS-IS Model) are reported in Table 1 and Table 2. The OS-IS model is a good benchmark to assess the improvement achieved with the ML method because its performance was assessed using the same instrumentation and the same data. Rarely the OS-IS model performed better than the others, except for the MAE of $T_p$. The reason is that the OS-IS model was designed to provide measures of $T_p$ only if $H_s$ was larger than 1 m, when the signal to noise ratio was higher. To make a fair comparison, the last columns of Table 1 and Table 2 report the validation metrics of a ML model (Boosting 20 pred.) trained and validated using the same dataset as for the OS-IS model (i.e., featuring $H_s>1$ m). As expected, its the validation metrics for $T_p$ and $T_m$ outperform those of the others, including those of the OS-IS model. This also highlights that the validation metrics depend on $H_s$ and, more in general, on the training/validation data.

| Algorithm | Random F. 20 pred. | Boosting 20 pred. | Boosting 15 pred. | Bagging 15 pred. | CNN Config. 1 | CNN Config. 2 | OS-IS Model | Boosting 20 pred. |
|---|---|---|---|---|---|---|---|---|
| Dataset | \multicolumn{7}{c}{Full Dataset} | | | | | | | $H_s>1$ m |
| RMS $H_s$ [m] | **0.17** | 0.18 | 0.18 | 0.17 | 0.19 | 0.23 | 0.21 | 0.24 |
| MAE $H_s$ [m] | **0.11** | 0.12 | 0.12 | 0.11 | 0.14 | 0.15 | 0.14 | 0.17 |
| Corr $H_s$ | **0.95** | 0.95 | 0.95 | 0.95 | 0.94 | 0.93 | 0.93 | 0.89 |
| RMS $T_p$ [s] | **1.38** | 1.38 | 1.38 | 1.38 | 1.40 | 1.47 | 1.52* | 0.97 |
| MAE $T_p$ [s] | 0.98 | 1.01 | 1.01 | 1.00 | 1.03 | 1.09 | 0.96* | 0.68 |
| Corr $T_p$ | **0.72** | 0.72 | 0.72 | 0.72 | 0.71 | 0.67 | 0.56* | 0.78 |
| RMS $T_m$ [s] | **0.63** | 0.63 | 0.64 | 0.65 | 0.70 | 0.66 | 0.95 | 0.57 |
| MAE $T_m$ [s] | **0.44** | 0.45 | 0.46 | 0.46 | 0.52 | 0.48 | 0.67 | 0.41 |
| Corr $T_m$ | **0.83** | 0.83 | 0.82 | 0.82 | 0.81 | 0.82 | 0.65 | 0.82 |

*Table 1. Performance of the ML models trained and validated over the data from 2014. Notice that, the ML model on the last column was trained only with data satisfying the condition $H_s>1$ m. The best performances are in bold. *The validation of OS-IS Model over $T_p$ was performed only for $H_s>1$ m.*

| Algorithm | Random F. 20 pred. | Boosting 20 pred. | Bagging 15 pred. | CNN Config. 2 | OS-IS Model | Boosting 20 pred. |
|---|---|---|---|---|---|---|
| Dataset | \multicolumn{5}{c}{November – December 2014} | | | | | $H_s>1$ m |
| RMS $H_s$ [m] | **0.19** | 0.20 | 0.20 | 0.23 | 0.24 | 0.23 |
| MAE $H_s$ [m] | **0.13** | 0.13 | 0.14 | 0.16 | 0.17 | 0.15 |
| Corr $H_s$ | **0.95** | 0.95 | 0.95 | 0.92 | 0.93 | 0.95 |
| RMS $T_p$ [s] | **1.28** | 1.30 | 1.27 | 1.32 | 1.47* | 1.10 |
| MAE $T_p$ [s] | **0.95** | 0.98 | 0.96 | 1.00 | 0.98* | 0.80 |
| Corr $T_p$ | **0.78** | 0.77 | 0.78 | 0.76 | 0.65* | 0.80 |
| RMS $T_m$ [s] | 0.83 | **0.79** | 0.84 | 0.80 | 1.13 | 0.73 |
| MAE $T_m$ [s] | 0.60 | **0.58** | 0.62 | 0.61 | 0.82 | 0.52 |
| Corr $T_m$ | 0.80 | **0.82** | 0.79 | 0.81 | 0.63 | 0.81 |

*Table 2. Performance of the ML models trained over the data from 2014 and validated over the data from November to December 2014 (6229 records). Notice that, the ML model on the last column was trained only with data satisfying the condition $H_s>1$ m. The best performances are in bold. *The validation of OS-IS over $T_p$ was performed only for data with $H_s>1$ m.*

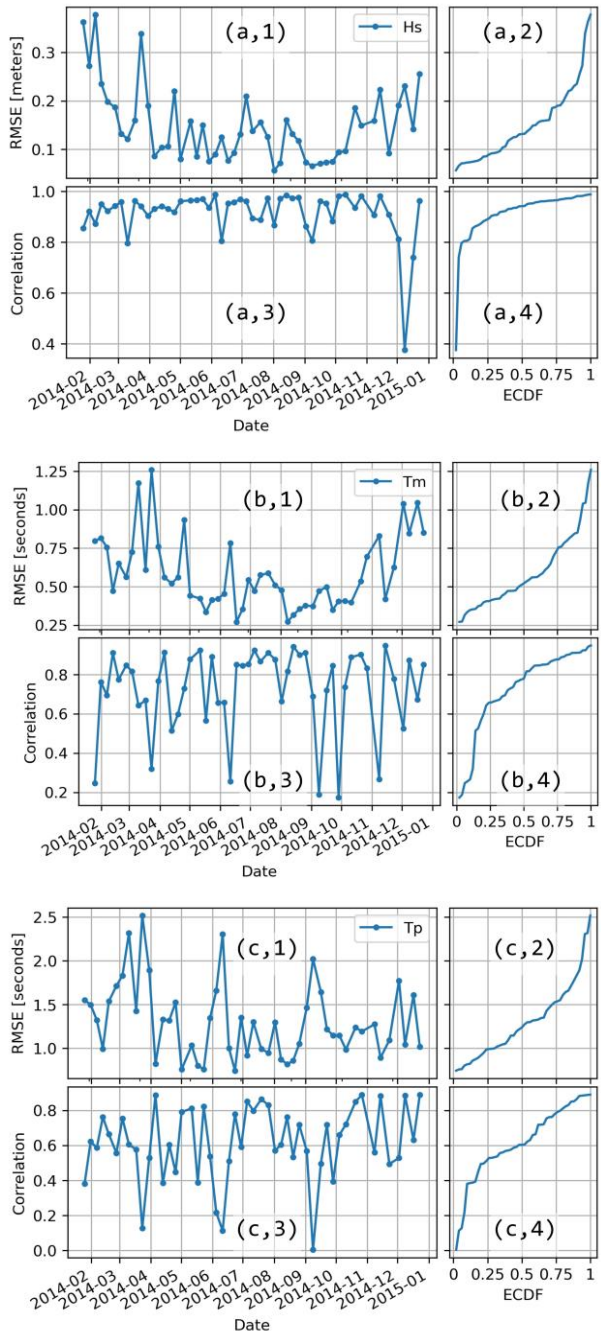

*Figure 11. Validation metrics (RMSE and correlation) of the Random Forest ML model (20 predictors) computed over 50 different time spans. a, b, and c labels are referred to $H_s$, $T_m$, and $T_p$ respectively. 1 and 3 are the time plots of RMSE and correlation respectively, whereas 2 and 4 are their ECDF plots.*

Comparing the ML models reported in Table 1, we notice that the difference between them is rather small, anyway, Random F. performed slightly better than the others on all metrics.

In Figure 11, the validation metrics (RMSE and correlation) are estimated 50 times over consecutive time intervals of about one week each (see "ML models validation" in Section 3.4). Notice the high variability of the estimates. On the right side of each time plot, the ECDF plots are shown and are useful to assess the confidence intervals. For example, in Figure 11 (a, 2) the median (ECDF=0.5) of the RMSE estimates is ~0.13 m and 90% (ECDF=0.9) of the RMSE estimates are smaller than ~0.25 m. Similarly, only 10% (ECDF=0.1) of correlation estimates are smaller than ~0.8.

Table 2 reports the validation metrics for a subset comprising only data from November to December 2014 (6229 records). Compared to the estimations in Table 1, the validation metrics are similar for $H_s$ and $T_p$, but they are slightly worse for $T_m$. The same validation data are shown in Figure 12 and Suppl. Figures 5 and 6. Specifically in these figures, the predicted values of $H_s$, $T_p$, and $T_m$ are compared to the desired outputs provided by the buoy. Inspecting these figures provides us with more insights than the validation metrics. For example, we notice that the absolute errors of $H_s$ (Figure 12 (b)), might be bigger than 1 meter (e.g., see the orange ellipse), but often this is due to a time-lag between the buoy measurements and predicted values. In fact, in subplot (a) we do not see a deviation bigger than 1 m between the buoy and predicted data (see the orange circle). Consequently, the MAE and the RMSE of $H_s$ reported in the tables might be overestimated. On the other hand, histograms and ECDF plots do not suffer from the time-lag issue. For example, Figure 12 (c) and (d) show a good agreement between buoy and predicted data. On the opposite, histograms and ECDF plots (c) and (d) of $T_p$ and $T_m$ in Suppl. Figure 5 and 6 do not overlap as nicely as with $H_s$. This indicates that assessing accurately $T_p$ and $T_m$ is slightly more complex than assessing the $H_s$.

Inspecting the errors histograms (subplots (e)) is a good way to evaluate if the ML model is introducing a systematic error. In fact, this would skew the histogram or displace its center away from zero; however, such effects were not detected in our case.

Finally, Figure 13 shows the hexagonal binning plots of the predicted values vs the desired values from the full dataset. Each plot also shows a black straight line along which the points are expected to align in the ideal case. We notice that for $H_s$ (subplot (a)), the points are mainly concentrated next to the black line for values smaller than 2 m whereas they are more spread for bigger values. However, the hue also indicates that values bigger than 2 m are much less frequent; consequently, the ML algorithm is less exposed to those values during the training, leading to worse accuracy. The spreading of $T_m$ and $T_p$, subplots (b) and (c), is not significantly larger for bigger values. However, we noticed that the points of $T_p$ (subplot (c)) do not distribute symmetrically about the black straight line, particularly for values between 2 and 4 seconds (inside the yellow circle). Most likely, this is driven by those records whose $H_s<1$ m. In fact, when the signal to noise ratio is lower (i.e., $H_s<1$ m), it is harder to identify the peak frequency (i.e., the peak period) than when the signal to noise ratio is higher (i.e., $H_s>1$ m). This agrees with the validation metrics reported in the last columns of Table 1 and Table 2.

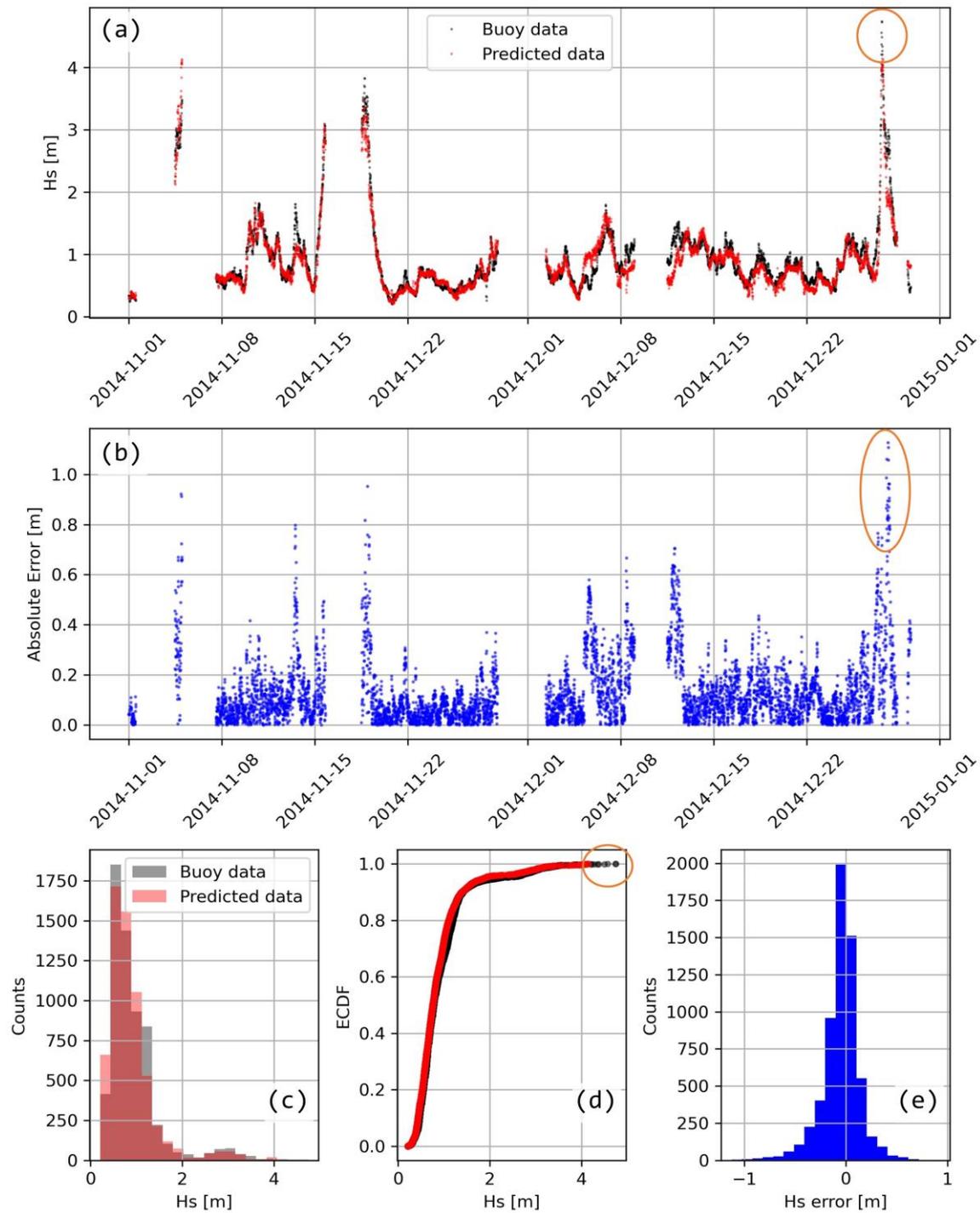

*Figure 12. Significant wave height ($H_s$) measured by the buoy (in black) and predicted by the Random Forest model (20 pred.) (in red) displayed in the time plot (a), the histogram (c), and the ECDF plot (d). The deviation between the two timeseries is displayed (in blue) in the time plot (b), and the histogram (e).*

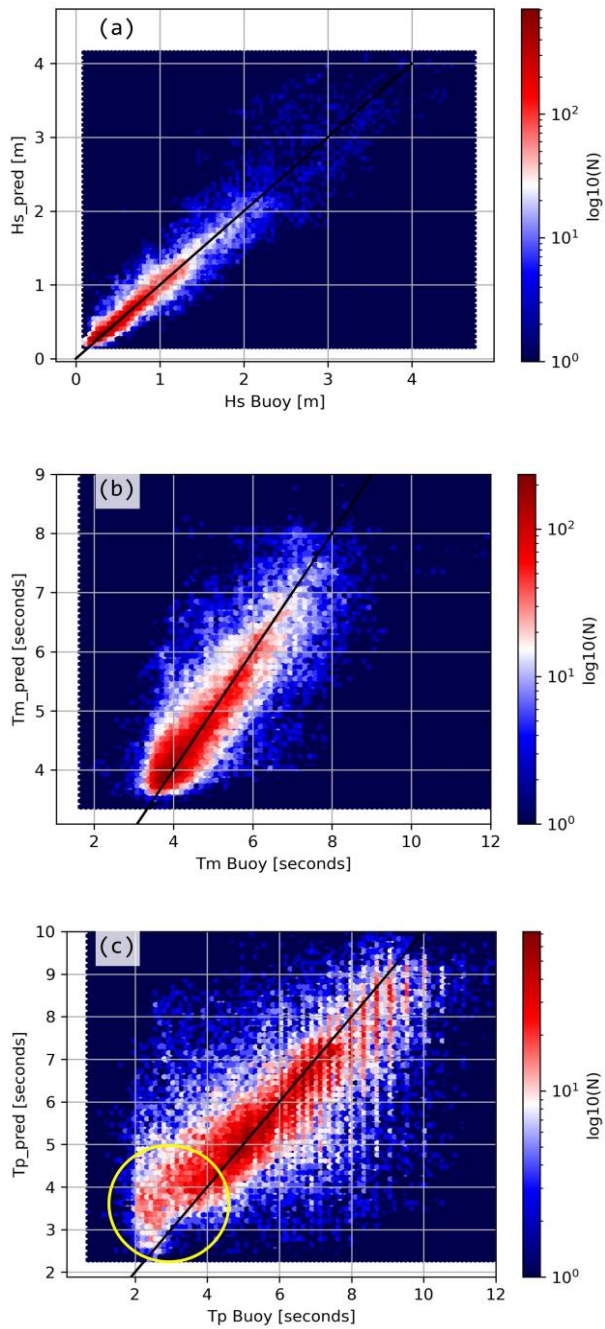

*Figure 13. Hexagonal binning plots of the $H_s$ (a), $T_m$ (b), and $T_p$ (c) values predicted by a Random Forest (20 predictors) model vs those measured by the buoy. Validation data from 2014.*

# 5 Discussion

The objective of this work has been to develop a sea wave monitoring system consisting of a moored buoy, a micro-seismic station, and a ML algorithm to automatically reconstruct missing buoy data using micro-seismic data. In this section, we discuss many aspects that differentiate our work from previous ones in the same field.

To begin, reconstructing missing buoy data using micro-seismic data is a novel aspect as, in previous works, other types of input data were used. For example, offshore hindcast and wind information were used in (Vieira et al. 2020) and weather data of adjacent buoys and bathymetric data were used in (Jörges et al. 2021). Moreover, many works were limited to short term (less than 48 hours) predictions meaning that their models rely on recent observations of the buoy, e.g., (Agrawal and Deo 2002). Instead, the proposed system can make predictions for any time after the last buoy observation keeping the same level of accuracy. This is key as buoy data gaps might be several weeks long.

As reported in Section 1, several previous works investigated the methods to retrieve sea wave data from micro-seismic data. A key aspect of the present work is the application of ML to automatically find the -micro-seismic to sea waves relationship- directly from data, with no need to develop and calibrate complicated geophysical models. Furthermore, the more training data are collected during the operations, the more accurately the ML model can perform. For example, by routinely training the ML algorithm as soon as new data is available, new types of events will be incorporated in the ML model with no need of manually tweaking its parameters, e.g., as in (Cutroneo et al. 2021). Despite these many advantages, ML limits the investigation on the input-output relationship (i.e., on the geophysical phenomenon) because most of the ML algorithms are "black boxes". Although this would be limiting in projects whose objective is to investigate the geophysical phenomenon, e.g., (Ardhuin et al. 2011), it is not relevant for missing data reconstruction.

Many aspects of the system were defined accordingly to the objective. For example, the micro-seismic station used for this work was specifically installed close to the shoreline and in front of the buoy of La Spezia (see Figure 2). As demonstrated in (Moschella et al. 2020), this displacement of the instrumentation leads to more accurate buoy data reconstruction. In fact, undesired signals, such as those from remote seismic sources, are less relevant and the signal to noise ratio is higher than that of stations far away from the shoreline and the reference buoy. For example in (Ferretti et al. 2013), the micro-seismic stations were from previous projects with different purposes (e.g., seismic monitoring) and, due to their locations, the interpretation of the signals was more complex.

Pursuing our objective, it was also important to use buoy data to train the ML algorithm and not weather model data, e.g., as done in (Cannata et al. 2020). In fact, if the sea wave data from weather models were a ground truth, there would be no need to retrieve the same data from any monitoring system. On the opposite, sea wave monitoring data are critical to validate weather models and implement data ingestion techniques.

In Section 3, we have gone through the main steps of the data processing; these are the typical steps of a ML project. Data cleaning is very important because ML algorithms are very sensitive to outliers and noisy data: the proposed method to clean the micro-seismic data is quite standard but still very effective; furthermore, it can be used to check the quality of the results during the monitoring operations. Despite the importance, data cleaning methods were seldom discussed in previous works. Feature engineering typically impacts the overall performance of ML models. We proposed two methods: one is based on using PSD arrays, the other on spectrograms; the latter was specifically meant for CNNs. Then, we introduced some ML algorithms and assessed their performance. Random forest was the best performing while CNNs did not shine as expected. Notice that the method to estimate and validate sea wave periods ($T_p$ and $T_m$) from the micro-seismic data was discussed. This was often disregarded in previous works and the validation of waves period estimates was rarely presented. Still, the waves period is an important parameter to assess the energy and, consequently, the impact of the sea waves.

We have shown that the performance assessments, in particular the validation metrics, depend on the validation dataset. For example, the periods are better evaluated when $H_s$ is bigger than 1 m; this should not be surprising because the signal to noise ratio is higher. To avoid wrong performance assessments, we proposed to use cross-validation, which is rarely used in projects related to the calculation of sea wave data from micro-seismic signals. Furthermore, we proposed a method to evaluate the confidence interval of the validation metrics using their ECDFs.

Our results for the accuracy on $H_s$ estimates are comparable to those of other works. However, accuracy depends on many variables (e.g., validation dataset, measurement site, sea wave conditions, time interval length) that make such comparisons meaningless. Consequently, the accuracy estimations should be used only to assess the ability of the system to reconstruct the missing buoy data.

It is worth to stress that the proposed system is an improvement of an existing technology for sea state monitoring, i.e., that of moored buoys. Accordingly, the micro-seismic system is not meant to operate as a standalone device. Therefore, the buoy is expected to continuously provide training data, particularly of sporadic events such as rare storm events; once these are incorporated in the training dataset, the ML algorithm will learn how to reconstruct them from the micro-seismic data. On the opposite, if the buoy is dismissed at some point in time, some new sporadic events might occur and the ML algorithm would not be able to reconstruct them accurately.

Finally, we stress the value of data science methods that are poorly and seldomly used in studies related to the evaluation of sea wave data from micro-seismic signals. Besides the ML methods, we have shown methods to compare timeseries and distributions. These are key when comparing and validating complex data such as sea wave measurements. For example, conventional scatter plots were often shown, where points formed a monolithic cloud hiding most of information. Instead of scatter plots, we proposed to use hexagonal binning plots. The validation metrics, such as RMSE, MAE, and correlation, were typically reported but, although very useful, they rarely tell the whole story. Instead, tools such as the ECDF plots were rarely used despite they are more informative than validation metrics.

# 6 Conclusion

In this work, we introduced a novel sea state monitoring system able to automatically reconstruct missing buoy data using micro-seismic data and machine learning. Specifically, we presented the methods to process the data, develop and train the ML algorithms, and assess their accuracy. As a case of study, we used the data collected in 2014 from a buoy of the Italian Data Buoy Network and a micro-seismic station (OS-IS). We demonstrated that many ML algorithms were able to reconstruct $H_s$, $T_m$, and $T_p$. However, the best performing was a Random Forest algorithm, whose root mean squared errors of $H_s$, $T_m$, and $T_p$ were respectively 0.17 m, 0.63 s, and 1.38 s. When $H_s$ was bigger than 1 m, the accuracy of $T_m$ and $T_p$ improved to respectively 0.57 s and 0.97 s.

By collecting more data, particularly for rare storm events, and tweaking the ML algorithm architectures, we believe that the accuracy can further improve. Specifically, we believe that the full potential of the CNN and the spectrograms as input, was not fully exploited. To do it, spectrograms with larger time span should be used and the CNN models should be more complex (i.e., more parameters). However, this requires more computational power than that available for this work.

# Declarations

**Funding**:   This research was funded by Assist in Gravitation and Instrumentation srl (AGI). The open access publication of this work was funded by Istituto Nazionale di Geofisica e Vulcanologia. The OS-IS station used in this work was financed by the European Regional Development Fund within the project Vento, Porti e Mare. The wave buoy data used in this publication were from ISPRA (Istituto Superiore per la Protezione e la Ricerca Ambientale – Italy) and made available by the EMODnet Physics project (www.emodnet-physics.eu/map, funded by the European Commission Directorate General for Maritime Affairs and Fisheries).
**Conflicts of interest/Competing interests**: Not applicable.
**Acknowledgements**: The wave buoy data used in this publication were from ISPRA (Istituto Superiore per la Protezione e la Ricerca Ambientale – Italy) and made available by the EMODnet Physics project (www.emodnet-physics.eu/map, funded by the European Commission Directorate General for Maritime Affairs and Fisheries).
**Data access:** The seismic data used in this work are available upon request and at authors' discretion.
**Authors' contributions.**
   **Lorenzo Iafolla**: Conceptualization, Methodology, Software, Validation, Investigation, Writing - original draft, Writing - review & editing, Visualization.
   **Emiliano Fiorenza**: Validation, Investigation, Writing - review & editing, Supervision.
   **Massimo Chiappini**: Writing - review & editing, Supervision, Project administration, Funding acquisition.
   **Cosmo Carmisciano**: Conceptualization, Methodology, Validation, Investigation, Writing - review & editing, Supervision, Project administration, Funding acquisition.
   **Valerio Antonio Iafolla**: Conceptualization, Methodology, Validation, Investigation, Writing - review & editing, Supervision, Project administration, Funding acquisition.